# LIC and LID considerations in the design and implementation of the MEMS laser pointing mechanism for the EUSO UV laser altimeter

Enrico Bozzo[1], Thomas Burch[2], Alessandra Ciapponi*[3], Clemens Heese[3], Arno Hoogerwerf[4], Nicholas Lan[5], Andrii Neronov[1], Vincent Revol[2], Ross Stanley[4],

[1] ISDC, Department of Astronomy, University of Geneva, Ch. d'Ecogia 16, 1290, Versoix, Switzerland, [2] CSEM Alpnach, Untere Gründlistrasse 1, CH-6055 Alpnach Dorf, Switzerland, [3] ESA – ESTEC, Kepleraan 1, 2201 AZ Noordwijk, The Netherlands, [4] CSEM SA, Rue Jaquet-Droz 1, CH-2002 Neuchâtel, Switzerland, [5] H.E. Space Operation, Huygensstraat 44-2A, 2201 DK Noordwijk ZH, The Netherlands

## ABSTRACT

The EUSO (Extreme Universe Space Observatory) project is developing a new mission concept for the scientific research of Ultra High Energy Cosmic Rays (UHECRs) from space. The EUSO wide-field telescope will look down from space onto the Earth night sky to detect UV photons emitted from air showers generated by UHECRs in our atmosphere. In this article we concentrate on the mitigation strategies agreed so far, and in particular on the implementation of a careful early selection and testing of subsystem materials (including optics), design and interfaces of the subsystem and an optimization of the instrument operational concept.

## 1. INTRODUCTION

The goal of the Extreme Universe Space Observatory (EUSO) to be installed on the International Space Station (ISS) is to detect about $10^3$ ultra-high energy cosmic ray (UHECR) "events" and produce, for the first time, a high signal statistics all-sky map of distribution of arrival directions of the highest energy cosmic rays with angular resolution close to 1 degree. This angular resolution is comparable to the typical deflection angle of UHECR by the Galactic magnetic field for the protons arriving at high Galactic latitude. EUSO would thus allow us to study UHECR sources using the conventional astronomical approach, i.e. identifying single objects and measuring the energy distribution of the cosmic rays coming from them.

### 1.1 The mission

EUSO [1] will use the Earth's atmosphere as a calorimeter of particle physics detector, seeing the ultraviolet fluorescence produced by the air molecules excited by the UHECR induced extensive air shower (EAS). The EUSO ultraviolet telescope has an aperture of 2.5 m, a field of view of $60°$, and will be operating in the 300-400 nm wavelength range. Ultraviolet light will be focused by a system of three Fresnel lenses made from UV transparent polymers and detected by an array of multi-anode photomultipliers (MAPMT) comprising $3 \times 10^5$ pixels. The cascade of high-energy particles, EAS, propagates with a speed close to the speed of light through the air in a direction that coincides with that of the primary UHECR initiating the EAS. The fluorescence emission appears for a short period of time along the path of the EAS (about 100 μs from the top of the atmosphere to the ground). EUSO will continuously take images of a large region of the atmosphere of the size of about 400 km across on a time scale of 2.5 μs. The angular resolution of the images will be $0.1°$ which corresponds to ~700 m linear distance on the ground assuming the 400 km altitude of the ISS. An UHECR induced EAS will thus appear during ~100 μs as a fast-moving source on the focal surface of the EUSO telescope. The combined information on the direction of motion and of the speed of the source provides an estimate of the arrival direction of the primary UHECR particle, while the intensity of the fluorescent emission gives a measurement of the energy of the primary UHECR.

*Alessandra.Ciapponi@esa.int, phone +31 71 565 8938

As the intensity of the UV light is strongly affected by the transmittance and scattering characteristics of the atmosphere around the shower, EUSO will be equipped with a dedicated Atmospheric Monitoring (AM) system consisting of an infrared camera and a LIght Detection And Ranging (LIDAR) device ([2], [4]). The IR camera [3] will provide an overall view of the distribution and altitude of the so-called low-altitude clouds in the telescope field of view. Complementary LIDAR data will provide measurements of the scattering and absorption properties of the atmospheric layers with optical depth down to 0.15 between the EAS and the EUSO telescope. The combined IR and LIDAR information of the meteorological conditions around the EAS are expected to provide a reconstruction accuracy of about 30% for the energy of the primary UHECR.

## 1.2 The EUSO lidar

The LIDAR is composed of a transmitting and a receiving system. The transmission system comprises a Nd:YAG laser and a pointing system to steer the laser beam in the direction of the triggered EAS events. As the laser backscattered signal will be received by the EUSO telescope (working as the LIDAR receiver), the laser operational wavelength was chosen to be the third harmonic of the Nd:YAG laser, at $\lambda = 355$ nm. The laser is being developed at RIKEN (Japan) and will be part of the JAXA (Japanese Space Agency) contribution to the Mission. The pointing system is under development at the University of Geneva (UoG, Switzerland), in close collaboration with Centre Swiss d'Electromagnetisme et de Microtechnique (CSEM, Switzerland).

An overview of the pointing system is show in Figure 1. The collimated laser beam is steered in the required directions by the Mirror Control System to within ±30° in order to completely cover the EUSO field of view. The position of the mirror is controlled by the control electronics using a feedback loop provided by a dedicated position sensor. The laser system and the mirror system are controlled by the Laser & Pointing Control Board, which is connected to the EUSO mission data processor and receives the trigger information to command the laser shooting and the mirror steering. In the current design, the mirror is inclined by 45° with respect to the impinging laser beam and thus it should able to tilt mechanically by ±15 degrees in order to point the laser beam by a maximal off-axis angle of ±30 degrees.

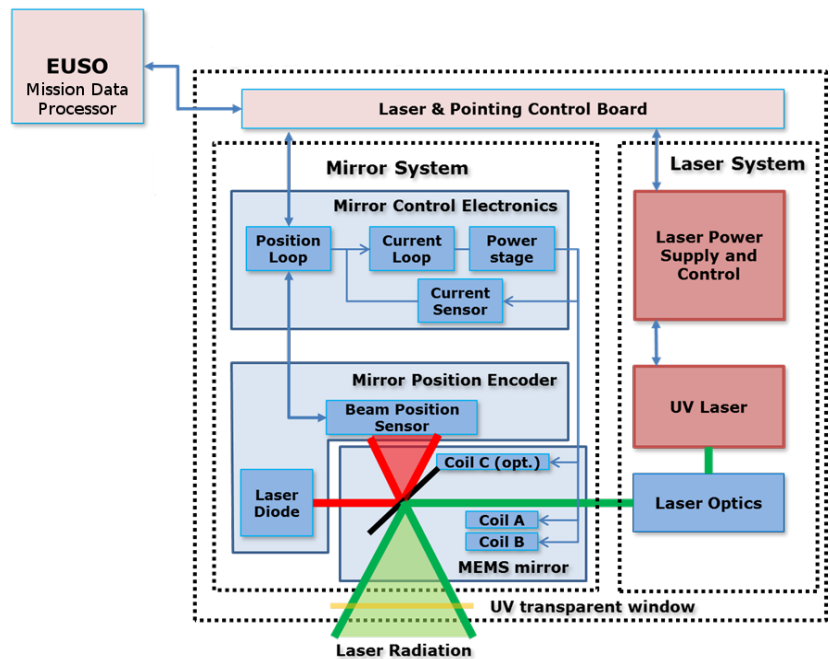

Figure 1. Overview of the LIDAR pointing system.

The interval between events that could be interpreted as Extensive Air Showers (EAS) is expected to be around 10 seconds. As a result, a new trigger for the LIDAR will be issued every 10 seconds on average. During this 10 seconds time interval between subsequent triggers, the LIDAR should be able to receive the information on the location of the last triggered event within the field-of-view, re-point the laser beam in this direction, and shoot several laser shots in the chosen direction (typically about 20 shots for each detected EAS candidate, see Sect. 2.2). The time needed to steer the laser beam is thus much shorter than the typical time between the triggering events, resulting in the requirement that the mirror should be able to move toward the EAS direction within a fraction of a second. This favours a lightweight mirror with little inertia and thus we presently expect to make use of an innovative Micro- Electro- Mechanical Systems (MEMS) pointing mechanism. This mechanism will be fabricated using microsystems technology and will use magnetic actuators to achieve the pointing function. It consists of a sandwich of two mechanical structures that allow a tip-tilt movement of the mirror that is squeezed in between. The bottom side of the mechanical structure can be forced in the X and Y directions independently by levers that are connected to it. The ends of these levers are connected to guided magnets that are actuated in the X and Y directions by coils (the current in the coils forces the magnets to move). The movement of these magnets is transferred by the levers that are rigid in the direction of the movement of the magnet and flexible in the orthogonal direction (i.e. around the second rotation axis). The movement is transferred by the levers to the bottom part of the mechanical structure. This movement results in a tip-tilt displacement of the mirror, since the top part of the structure is rigid in the X and Y directions but both are flexible in the tip and tilt rotations.

The combination of a large deflecting angle and a large mirror surface (about 3x3 mm) presents a non-negligible design challenge for the MEMS mirror. On one hand, the large mirror surface is necessary to have the possibility of spreading the impinging laser energy density over a reasonable area and keep its average value below the critical threshold of 0.5 J/cm$^2$ (See Sect. 2.1). On the other hand, the foreseen large deflection angles of ±15 degrees to be covered by the mechanisms, in combination with the above mentioned mirror surface, require a wide displacement of the mirror edges and thus a magnetic actuation force working over long distances. Simulations of the mirror deflection have been already carried out to prove the suitability of the present unit design. The simulations also showed that both a mechanical clamping and a magnetic damping mechanism needs to be implemented in order for the unit to be able to sustain launch shocks and achieve the required pointing stability from one laser shoot to the other, respectively. The current implementation of the clamping and damping mechanisms is shown on the left side Figure 2, together with an overall sketch of the entire unit.

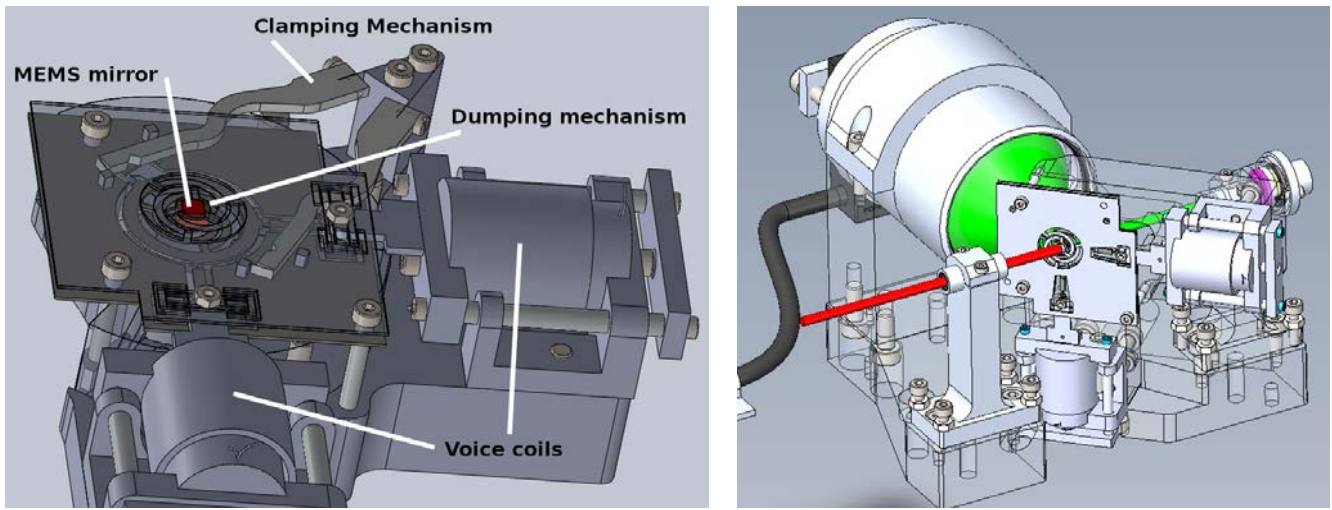

Figure 2: Left: the current design of the laser and pointing system EBB. All main components are shown, including the MEMS mirror, the voice coils, the clamping mechanism, and the damping mechanism. Right: a sketch of the current FUMO design. The red beam corresponds to the main EUSO laser beam, while the green one is the beam used for the closed loop with the optical sensor. The MEMS mirror and the voice coils are also visible. The clamping and damping mechanisms have been removed from the FUMO.

### 1.3 Project status in Switzerland

The preliminarily design of the EUSO Laser and Pointing system has been presented by CSEM at the end of the project phase B1, concluded in 2015. The scope of the currently on-going phase B2 is the manufacturing of a fully functional Elegant Bread-Board (EBB) of the unit, which is able to achieve the required performances but it does include mostly of the shelf components (i.e. not yet suitable for a full space qualification). As part of the EBB, the mirror control electronics and the laser and pointing system control board will also be developed. A particular attention is being paid to the issue of laser induced damages (LID) and contamination (LIC). For this reason, a number of dedicated tests of the MEMS sub-assembly are being planned for the current phase (see Sect. 2). CSEM is also developing well in advance of the final EBB delivery an intermediate functional model (FUMO), devoted to help software and algorithms development as well as improvements on the final EBB design through the execution of numerous functional tests. The FUMO includes all the key components of the EBB: the MEMS mirror, the voice coils, the optical mirror position sensor and the optical path of the UV Laser. The clamping and damping mechanisms are not included. A sketch of the FUMO is shown on the right side of Figure 2. The commissioning of the FUMO was completed in May 2016, and all foreseen tests will be executed within the summer of 2016.

## 2. LIC AND LID MITIGATION MEASURES IN THE SUBSYSTEM DESIGN

The space environment present particular challenges for optics that are not encountered in on ground operation: vacuum environment, outgassing, radiation, long life time without maintenance, etc. A de-risking campaign needs to take into account the specific environment where the instrument will operate and then begin the testing phase on the most representative conditions. The materials used in the system shall be contaminant free or complaint with the maximum contamination level that must be agreed for each mission. The interaction of the outgassing particles with the high UV energy beam will deposit a highly absorbing layer on the surface of the optics. Contamination can drastically accelerate the degradation of the optical performances until, eventually, a catastrophic damage will occurred.

Challenges in space-borne laser design are intensified at the shorter UV wavelengths, compared to the longer IR one, requiring more stringent precautions to be taken. As a result of ESA experience in the framework of UV laser altimeter developments (in particular for the ADM-AEOLUS and Earthcare missions), risk mitigation strategies have been agreed and implemented within the development plan of the MEMS pointing device to address issues related to the Laser Induced Contamination (LIC) and Laser Induced Damage (LID).

For example, screening of particular materials and a dedicated campaign for the choice of the optics supplier will start in the early phase of this mission [4, 5, 6, 7]. Conventional contamination control techniques have been found to be insufficient for space-borne lasers in the UV range due to particular material and process sensitivity. While mechanisms for LIC are still not completely understood, such early test campaigns have been found to be the most reliable method of risk mitigation.

### 2.1 Mirror specifications

The requirements for the mirrors of the MEMs system are quite challenging. In Table 1 a summary shows some of the most important parameters in terms of design constraints. The energy of the incoming laser has been limited to 5 mJ to avoid exceeding the maximum fluence at mirror level. Another important parameters is the lifetime: the mirrors will be irradiated with $2 \times 10^8$ pulses during the 3 years foreseen for the mission. Degradation of the performances could lead to a failure of the mission. The fact that the mirrors must work at different angles of incidence make the design of the coating stack very demanding.

| Criterion | Min | Typical | Max | Units/comments |
|---|---|---|---|---|
| Wavelength | | 355 | | nm |
| Substrate | | UV grade fused silica | | |
| Diameter | | 3.5 | | mm |
| Thickness | | 2 | | mm |
| Reflectivity (s-pol) | 98 | 99 | | % |
| Angle of incidence | 30 | 45 | 60 | degree |
| Flatness | | $\lambda/10$ | | @635nm |
| Scratch-dig | | $10^{-5}$ | | |
| Laser fluence | | 0.5 | | $J/cm^2$, already reduced from the 3 $J/cm^2$ foreseen at start of subsystem phase |
| Lifetime | | $2 \times 10^8$ | | Pulses at max fluence |

Table 1: Requirement for the mirror that will be used in the MEMS subsystem in order to reach the performances request by the mission.

### 2.2 LIC and LID mitigation strategy

Mitigation measures may be split into the following categories, having been discussed previously as to the most sensible course of action to take for this phase of the subsystem design. The main mitigation measures agreed in the frame of this phase of the subsystem development are summarised below.

1. Mitigation measures that involve subsystem external interfaces, and instrument operational concept. These measures must be agreed at instrument level, as well as with the relevant subsystems external to this laser pointing mechanism. However, as the current development is considered to be a standalone activity aimed at increasing the TRL (Technical Readiness Level) of the subsystem, the following main points have been taken as assumptions within the scope of this phase B2 to allow progresses. Agreements at instrument level and other subsystems will be pursued in parallel.
    a. Limiting the peak fluence incident on the laser pointing mechanism mirror to <0.5 $J/cm^2$ as a target value to mitigate LID. This would entail a flat rather than Gaussian incident beam profile and optimization of the instrument operating concept to utilise multiple 5mJ laser pulses (about 20 for each detected EAS event) rather than a lower number (5, as originally planned) of single 20 mJ pulses.
    b. To mitigate LIC, the subsystem is now envisaged to be enclosed in a compartment pressurised with ~1 bar technical air. The selection of the materials will follow the ECSS standard in terms of CVCM (Collected Volatile Condensable Materials) and TML (Total Mass Loss) [4]. In addition, where possible, materials (accounting for manufacturing processes) envisaged to be in the vicinity of these pointing mechanism optics will undergo careful preselection for intended use in flight and early testing within this phase to identify potential bad actors [9].
2. Mitigation measures involving the subsystem design and development
    a. Careful design and selection of laser pointing mechanism mirror and coatings. In addition to the difficulties with respect to operating this mechanism at 355nm, the large angle of incidence required for the mirror due to the large pointing range of the mechanism implies that mirror and coating designs are challenging. In particular, consideration of a mirror compliant with LIC and LID issues may imply an increase of the mirror dimensions. Such a change would necessitate some redesign of the MEMS mechanism. A process of careful mirror selection is in progress, involving early screening for LID behaviour of reference COTS HR355 mirror samples prior to selection of candidates for dedicated LID

b. Dedicated test campaigns aimed at de-risking the unit design with respect to LID and LIC. A dedicated breadboard (the LIB, i.e. laser induced breadboard) is envisaged within the current activity (in addition to the EBB and the FUMO, see Sect. 1.3) aimed at proving the function and performance of the MEMs pointing mechanism. The LIB will not be actively actuated but is envisaged to contain a mirror customised to EUSO requirements along with material samples deemed to be representative of items that will be in the vicinity of these optics in flight. This breadboard is intended to be enclosed and pressurised with ~1 bar technical air.
      i. LIC – This dedicated breadboard is envisaged to undergo clean material allowing testing to be adequately representative of flight conditions. By including material samples (as representative as possible in relation to envisaged flight design) in the vicinity of the optics, a test campaign using this breadboard will utilise representative laser conditions (e.g. number of shots, fluence, repetition rate, etc), to identify potential issues in terms of LIC.
      ii. LID – The same breadboard described above is intended to undergo a combined test campaign to address both LIC and LID. In terms of LID specifically, the test campaign will subject the chosen optics to representative but realistic worst case conditions, in particular with respect to number of shots of the intended instrument lifetime and the target peak fluence of <0.5 J/cm$^2$.

and LIC test campaigns. The latter will make use of mirrors procured according to the specific EUSO requirements.

As a first action, several optical suppliers have been screened to select the best candidate(s) that will be chosen for the flight optics. Two standard mirrors 45 degree, HR at 355 nm from the different suppliers have been tested following a S-on-1 test procedure (ISO 21254-2) in laboratory environment conditions. The results will be used to write a short list of potential suppliers that will be then contacted with the specific requirement sets for the EUSO mirror (Table 1). The next step will be the tests in representative environmental condition with a 10k-on-1 procedure, followed by a raster scan at nominal fluence for an area as large as the actual beam spot. An endurance test shall also be performed to confirm performances for a long period of irradiation [9].

## 3. CONCLUSION

The development of the LIDAR for the EUSO mission is currently on-going at the University of Geneva in collaboration with CSEM (Switzerland) and supported by the ESA PRODEX office. Several activities and investigations are being focused in the present B2 phase of the project on the LID and LIC, considered to be potential issues for the EUSO LIDAR (given its extended lifetime and operations in the UV domain). Careful selection of the materials, pre-screening with ad-hoc tests, fluence limitation, etc are some of the lesson learned from ESA previous experiments with UV lasers in space that will be applied to the EUSO LIDAR development to ensure a higher likelihood of success. An early de-risking campaign will benefit the entire mission and will give a high confident level for the implementation of the further phases.

## REFERENCES

[1] Adams, J. H., et al. 2015, ExA, 40, 3.
[2] Adams, J. H., et al. 2015, ExA, 40, 45.
[3] Adams, J. H., et al. 2015, ExA, 40, 61.
[4] Rodríguez Frías, M. D., Toscano, S., Bozzo, E., del Peral, L., Neronov, A., Wada, S., EPJWC, 8902007, 2015.
[5] ECSS-Q-ST-70-01C, "Space Product Assurance: Cleanliness and contamination control", November 2008.
[6] Riede, W., Schroeder, H., Bataviciute, G., Wernham, D., Tighe, A., Pettazzi, F., and Alves, J., "Laser-induced contamination on space optics", *Proc. SPIE* 8190, Laser-Induced Damage in Optical Materials: 2011, 81901E (2011).

[7] Schroeder, H., Wagner, P., Kokkinos, D., Riede, W., and Tighe, A.," Laser-induced contamination and its impact on laser damage threshold ", *Proc. SPIE* 8885, Laser-Induced Damage in Optical Materials: 2013, 88850R (2013).

[8] Burnham, A., Runkel, M., Demos, S.G., Kozlowski, M.R., Wegner, P.J., "Effect of vacuum on the occurrence of UV-induced surface photoluminescence, transmission loss, and catastrophic surface damage", Photonics for space environments VII, *Proc. SPIE* 4134, 243-252 (2000).

[9] Riede, W., Allenspacher, P., Lammers, M., Wernham, D., Ciapponi, A., Heese, C., Jensen, L., Maedebach, H., Schrameyer, S., and Ristau, D., "From ground to space: how to increase the confidence level in your flight optics", *Proc. SPIE* 8885, Laser-Induced Damage in Optical Materials: 2013, 88850D (2013).